\documentclass[11pt]{article}
\usepackage{color}
\usepackage{bm}
\usepackage{slashed}
\usepackage{varioref}
\usepackage{xcolor} 
\usepackage{graphicx}
\usepackage{amsmath}
\usepackage{amssymb}
\usepackage{amsfonts}
\usepackage{amsthm}
\usepackage{array}

\usepackage[
      colorlinks=true,
      linkcolor=blue,
      urlcolor=blue,
      filecolor=blue,
      citecolor=red,
      pdfstartview=FitV,
      pdftitle={},
      pdfauthor={},
      pdfsubject={},
      pdfkeywords={},
      pdfpagemode=None,
      bookmarksopen=true
]{hyperref}

\usepackage{epsfig}

\usepackage{hyperref}
\def\half{\frac{1}{2}}

\def\beq{\begin{eqnarray}}
\def\eeq{\end{eqnarray}}

\def\w{\wedge}
\def\be{\begin{equation}}
\def\ee{\end{equation}}
\def\bea{\begin{eqnarray}}
\def\eea{\end{eqnarray}}

\def\be{\begin{equation}}
\def\ee{\end{equation}}
\def\bea{\begin{eqnarray}}
\def\eea{\end{eqnarray}}

\labelformat{equation}{Eq.~(#1)} 
\labelformat{appendix}{Appendix #1}

\newcommand{\rom}[1]{\mathrm{#1}}

\def\cE{\mathcal{E}}
\def\cF{\mathcal{F}}

\def\cK{\mathcal{K}}

\def\cS{\mathcal{S}}
\def\cT{\mathcal{T}}

\def\nn{\nonumber}

\newcommand\Tau{\mathcal{T}}

\newcounter{mnotecount}[section]

\renewcommand{\themnotecount}{\thesection.\arabic{mnotecount}}

\newcommand{\mnotex}[1]
{\protect{\stepcounter{mnotecount}}$^{\mbox{\footnotesize
$
\bullet$\themnotecount}}$ \marginpar{
\raggedright\tiny\em
$\!\!\!\!\!\!\,\bullet$\themnotecount: #1} }

\date{}
\numberwithin{equation}{section}

\begin{document}

\title{\bf First law of black hole mechanics with fermions}

\author{P. B. Aneesh\footnote{aneeshpb@cmi.ac.in}$~^{1}$, Sumanta Chakraborty\footnote{tpsc@iacs.res.in}$~^{2,3}$, \\Sk Jahanur Hoque\footnote{skjhoque@cmi.ac.in}$~^{1,4}$ and Amitabh Virmani\footnote{avirmani@cmi.ac.in}$~^{1}$
\vspace{0.6cm} \\
$^{1}${\small{Chennai Mathematical Institute,}} 
\\ {\small{H1 SIPCOT IT Park, Kelambakkam, Tamil Nadu 603103, India }}
\vspace{0.3cm} \\
$^{2}${\small{School of Mathematical and Computational Sciences}}
\\
{\small{Indian Association for Cultivation of Science, Kolkata 70032, India}}
\vspace{0.3cm} \\
$^{3}${\small{School of Physical Sciences}}
\\
{\small{Indian Association for Cultivation of Science, Kolkata 70032, India}}
\vspace{0.3cm} \\
$^{4}${\small{Institute of Theoretical Physics,
Faculty of Mathematics and Physics, Charles University,}}
\\
{\small{V~Hole\v{s}ovi\v{c}k\'ach 2, 180~00 Prague 8, Czech Republic}}
}
\maketitle 
\begin{abstract}
In the last few years, there has been significant interest in understanding the stationary comparison version of the first law of black hole mechanics in the vielbein formulation of gravity. Several authors have pointed out that to discuss the first law in the vielbein formulation one must extend the Iyer-Wald Noether charge formalism appropriately. Jacobson and Mohd [arXiv:1507.01054] and Prabhu [arXiv:1511.00388]  formulated such a generalisation for symmetry under combined spacetime diffeomorphisms and local Lorentz transformations. In this paper, we apply and appropriately adapt their formalism to four-dimensional gravity coupled to a Majorana field and to a Rarita-Schwinger field. We explore the first law of black hole mechanics and the construction of the  Lorentz-diffeomorphism Noether charges in the presence of fermionic fields, relevant for simple supergravity.
\end{abstract}
\newpage

\tableofcontents
\section{Introduction}

 Iyer and Wald \cite{Iyer:1994ys} gave a derivation of the
stationary comparison version of the first 
law of black hole mechanics for arbitrary perturbations around a stationary axisymmetric black hole in any diffeomorphism covariant theory of gravity. The black hole entropy was identified with the integral over the bifurcation surface of the diffeomorphism Noether charge for the horizon generating Killing field.  The results of Iyer and Wald have found numerous applications over the years~\cite{Hollands:2005wt, Hollands:2012sf, Faulkner:2013ica}.

However, as emphasised in detail in references~\cite{Jacobson:2015uqa, Prabhu:2015vua}, there are situations of physical interest where Iyer-Wald analysis cannot be  applied directly. An assumption that goes into their analysis is that matter fields, if present, are smooth tensor fields on the spacetime. Often in gauge theories, one cannot always make a gauge choice such that the gauge fields are smooth tensor fields. A similar situation arises for gravity theories  written in vielbein formulation~\cite{Jacobson:2015uqa}, where the spin-connection might not be smooth in some chosen gauge (i.e., for a choice of vielbeins $e_\mu^a$). Since the coupling of fermions to gravity is typically through the vielbeins, Iyer-Wald analysis cannot be directly applied to gravity coupled to fermions. In the last few years, there has been interest in understanding the stationary comparison version of the first law of black hole mechanics in the vielbein formulation and renewed interest in the theory of Noether charges in the vielbein formulation of gravity, see e.g.,~\cite{Randono:2009ci, 
Barnich:2016rwk, Frodden:2017qwh, DePaoli:2018erh, Oliveri:2019gvm}.   

In order to discuss the first law in the vielbein formulation, one must extend the Iyer-Wald's Noether charge formalism
 appropriately. 
  Prabhu~\cite{Prabhu:2015vua} formulated a generalisation for symmetry under combined diffeomorphisms and internal gauge transformations in terms of fields living on a principal bundle over spacetime. Using this formalism he presented a derivation of the first law of black hole mechanics in the vielbein formulation of gravity, for gravity coupled to  Yang-Mills field and for gravity coupled to a Dirac fermion. Interestingly, he found that the contribution due to the  Dirac field to the Lorentz-diffeomorphism\footnote{A Lorentz-diffeomorphism is a diffeomorphism accompanied by a Lorentz transformation, i.e., a frame rotation. The particular Lorentz transformation of the frames, given a diffeomorphism, is given by Kosmann derivative introduced in \ref{Kosmann_vielbein}.}  Noether charge vanishes on-shell for the horizon generating Killing field at the bifurcation surface.    
Thus, we are led to the question: does the contribution to the Lorentz-diffeomorphism Noether charge  vanish generically for fermions at the bifurcation surface?

 The situation should be compared to bosonic fields. For a minimally coupled scalar too the contribution to the diffeomorphism Noether charge due to the scalar vanishes. Though, this is not the case for a vector field (see e.g., \cite{Gao:2003ys}). Thus, it is interesting to explore the case of the Rarita-Schwinger field, which besides being fermionic also carries a spacetime index. In this paper, we apply and appropriately  adapt the formalism of~\cite{Jacobson:2015uqa, Prabhu:2015vua} to four-dimensional gravity coupled to a Majorana field and to  a Rarita-Schwinger field. We explore the first law of black hole mechanics and the construction of  Lorentz-diffeomorphism Noether charges.  With the Rarita-Schwinger field one can in principle write down a few different diffeomorphism covariant Lagrangians. Perhaps the most natural set-up to consider is the case of $N=1, D=4$ supergravity, often called simple supergravity.

 At this stage, it is important to note that local supersymmetry is not an  internal gauge symmetry; it is a spacetime symmetry. Prabhu's formalism \cite{Prabhu:2015vua}, although quite general, is restricted to the cases of internal gauge symmetry. Thus, local supersymmetry cannot be properly taken into account within the framework of references~\cite{Jacobson:2015uqa, Prabhu:2015vua}. For this reason, we `switch off' supersymmetry in our  study of the Lorentz-diffeomorphism Noether charges. Perhaps an independent formulation of the first law can be achieved by considering supergravity as a geometric theory on superspace, where the fermionic gauge symmetry is properly taken into account. We leave this for future work.

The rest of the paper is organised as follows. In section \ref{sec:summary} we summarise the Lorentz diffeomorphism Noether charge formalism. For simple supergravity, the Rarita-Schwinger field is a Majorana vector-spinor. The Majorana condition brings in new elements in the computation. It is instructive to study the simpler spin-1/2 Majorana field first, before diving into the case of the simple supergravity. Therefore, in section \ref{sec:Majorana} we study the Lorentz-diffeomorphism Noether charge for the simpler case of spin-1/2 Majorana field. This section connects to the case of the Dirac field studied in reference~\cite{Prabhu:2015vua}. It also serves as a warm-up for simple supergravity considered in section \ref{sec:supergravity}. The key result of section \ref{sec:supergravity} is an expression for the contribution to the Lorentz-diffeomorphism Noether charge due to the Rarita-Schwinger field. Using this Noether charge we formulate a stationary comparison version of the first law in section  \ref{sec:first_law}. We close with a brief discussion in section \ref{sec:discussion}.

\section{A summary of the Lorentz-Diffeomorphism Noether charge formalism}\label{sec:summary}

For theories with internal gauge symmetries there is no natural action of spacetime diffeomorphisms on dynamical fields. We only have a notion of spacetime diffeomorphisms upto gauge transformations. Without such a separation between spacetime diffeomorphisms and gauge transformations, the \emph{diffeomorphism} Noether charge is not an adequate concept to work with. This issue has been discussed in the context of the first law of black hole mechanics over the years~\cite{Sudarsky:1992ty, Sudarsky:1993kh, Gao:2003ys, Corichi:2010ur, Corichi:2013zza} and, from our point of view,  is satisfactorily addressed in \cite{Jacobson:2015uqa, Prabhu:2015vua}.   A different perspective on these issues is presented in \cite{DePaoli:2018erh}.

Prabhu \cite{Prabhu:2015vua} formulates a given gravity theory in terms of fields living on a principal bundle over spacetime. Then he considers the full group of transformations, diffeomorphisms together with gauge transformations, viewed as automorphisms of the bundle. For a general such  automorphism he defines a notion of Noether charge and uses it to obtain a first law of black hole mechanics as a variational identity. A key idea in this construction is the fact that the variation of the fields under a  general automorphism $X$  is simply given by  \emph{the  standard Lie derivative} with respect to vector field $X$ \emph{on the bundle}.

Jacobson and Mohd in reference~\cite{Jacobson:2015uqa} take a more pragmatic approach. They propose a generalisation of the Iyer-Wald diffeomorphism Noether charge for a given spacetime diffeomorphism to what they call \emph{Lorentz-Diffeomorphism}Noether charge. The two approaches have some similarities and some differences. In general there is no unique way to associate an automorphism of the bundle to a given spacetime diffeomorphism. However, the requirement that the vielbeins (co-frames) be preserved by the corresponding Killing vector of a given spacetime metric, uniquely determines the infinitesimal bundle automorphism. A proof of this statement with references to the original literature can be found in \cite{Jacobson:2015uqa, Prabhu:2015vua}. This uplift defines a Lorentz-Lie derivative. We use the notation $\mathcal{K}$ following  \cite{Jacobson:2015uqa} where the Lorentz-Lie derivative was defined for arbitrary spacetime vector fields by the \emph{same formula} as for the Killing vector fields. We call it the Kosmann derivative.  On co-frames, for \emph{arbitrary} $\xi$,
\be
\mathcal{K}_\xi e_\mu^a = \mathcal{L}_\xi e_\mu^a + \left( E^{\nu[a} \mathcal{L}_\xi e_\nu^{b]} \right) e_{b\mu}, \label{Kosmann_vielbein}
\ee
where  $\mathcal{L}_\xi $ is the standard Lie derivative computed by ignoring the internal indices, and $E^\mu_a$ are the inverse vielbeins. 
Since in the later parts of the paper we suppress spacetime indices and work in the form notation, a separate notation is required for the vielbeins and inverse vielbeins. The vielbeins and inverse vielbeins satisfy the usual properties, 
$E_a \, \cdot \, e^b := E_a^\mu e^b_\mu = \delta_a^b$, and under arbitrary variation $\delta E_a \cdot e^b = - E_a \cdot \delta e^b$. 

In order to generalise the concept of Kosmann derivative to an arbitrary tensor-spinor object carrying both spacetime and/or Lorentz/spinor indices, we first rewrite  \ref{Kosmann_vielbein} as:
\bea
\mathcal{K}_\xi e_\mu^a &=& \mathcal{L}_\xi e_\mu^a + \frac{1}{2} \left(
\xi \cdot  \mathring{\omega}_{cd} +
E^\rho_c E^\sigma_d ~\partial_{[\rho}\xi_{\sigma]} \right) \left( 2  \eta^{a[c} \delta^{d]}{}_b \right) e^b_\mu, \\
&=& \mathcal{L}_\xi e_\mu^a + \frac{1}{2} \left(
\xi \cdot  \mathring{\omega}_{cd} +
\mathring{D}_{[c}\xi_{d]} \right) \left( 2  \eta^{a[c} \delta^{d]}{}_b \right) e^b_\mu,
\eea
where $\mathring{\omega}_{cd}$ is the  torsionless part of the spin-connection and $\mathring{D}_\mu$ is the torsionless Lorentz covariant derivative.  This rewriting makes the vector representation of the Lorentz group in equation  \ref{Kosmann_vielbein} manifest, and allows us to propose a Kosmann derivative for an arbitrary tensor-spinor object~\cite{Ortin:2002qb}. For \emph{an arbitrary spacetime vector field} $\xi$, we define,
\bea
\mathcal{K}_{\xi}T_{\mu_{1}\cdots\mu_{m}}{}^{\nu_{1}\cdots\nu_{n}}
& \equiv & 
\mathcal{L}_\xi T_{\mu_{1}\cdots\mu_{m}}{}^{\nu_{1}\cdots\nu_{n}} + 
\frac{1}{2} \left( \xi \cdot \mathring{\omega}_{ab} + \mathring{D}_{[a}\xi_{b]} \right) \Gamma_{r}(M^{ab})
T_{\mu_{1}\cdots\mu_{m}}{}^{\nu_{1}\cdots\nu_{n}},
\label{Kosmann_general}
\eea
where the Lorentz/spinor indices are suppressed and the Lie derivative is the usual Lie derivative that only sees the spacetime indices, and where $\Gamma_{r}(M^{ab} )$ are the representation matrices for the Lorentz generators $M^{ab}$ in the representation $r$ of the Lorentz tensor-spinor $T$. 
For vector representation, 
\be
\Gamma_\rom{vec}(M^{ab})^c{}_{d} =   2  \eta^{c[a} \delta^{b]}{}_d, 
\ee
and for the four-dimensional spinor representation, 
\be
\Gamma_\rom{spinor}(M^{ab}) =   \frac{1}{2} \gamma^{ab} =  \frac{1}{4} \left(\gamma^{a} \gamma^{b} - \gamma^{b} \gamma^{a}\right). 
\ee
For spinors, gamma matrices, Majorana condition, Majorana flip conventions we exclusively follow reference~\cite{Freedman:2012zz}.

It then follows that for spinors, the Kosmann derivative takes the form,
\be
\mathcal{K}_\xi \Psi = \xi^\mu  \mathring{D}_\mu \Psi  + \frac{1}{4} \partial_{[\mu} \xi_{\nu]} \gamma^{\mu \nu} \Psi, \label{Kosmann_spinor}
\ee
where $\mathring{D}_\mu$ is the torsionless spinor covariant derivative,
\be
\mathring{D}_\mu \Psi (x) = \left( \partial_\mu + \frac{1}{4} \mathring{\omega}_{\mu ab} \gamma^{ab} \right) \Psi(x),
\ee
and where $\mathring{\omega}_{\mu}{}^{ab} $ and  the  Christoffel symbol  $\mathring{\Gamma}^\sigma_{\mu \nu}$ are related by the vielbein postulate, 
\be
\partial_\mu e^a_\nu + \mathring{\omega}_\mu{}^a{}_{b} e^b_\nu - \mathring{\Gamma}^\sigma_{\mu \nu} e_\sigma^a =0.
\ee

Along identical lines, it also follows that the Kosmann derivative for the Rarita-Schwinger field, which is a vector-spinor field, takes the form,
\bea
\mathcal{K}_{\xi} \psi_{\mu}&=&\mathcal{L}_{\xi} \psi_{\mu}+\frac{1}{4} \left(\xi \cdot \mathring{\omega}_{ab} +\mathring{D}_{[a}\xi_{b]} \right)\gamma^{ab}\psi_{\mu}
\nonumber
\\
&=&\xi^{\alpha}\partial_{\alpha}\psi_{\mu}+\psi_{\alpha}\partial_{\mu}\xi^{\alpha}+\frac{1}{4} \left(\xi \cdot \mathring{\omega}_{ab} +\mathring{D}_{[a}\xi_{b]} \right)\gamma^{ab}\psi_{\mu}
\nonumber
\\
&=&\xi^{\alpha}\left(\partial_{\alpha}\psi_{\mu}-\partial_{\mu}\psi_{\alpha}+\frac{1}{4}\mathring{\omega}_{\alpha}^{~~ab}\gamma_{ab} \psi_{\mu}
-\frac{1}{4}\mathring{\omega}_{\mu}^{~~ab}\gamma_{ab} \psi_{\alpha}\right)+\partial_{\mu}\left(\xi^{\alpha}\psi_{\alpha}\right)
\nonumber
\\
&&+\frac{1}{4}\mathring{\omega}_{\mu}^{~~ab}\gamma_{ab} \left(\xi^{\alpha}\psi_{\alpha}\right)
+\frac{1}{4}\mathring{D}_{[a}\xi_{b]}\gamma^{ab}\psi_{\mu}
\nonumber
\\
&=&\xi^{\alpha}\left(\mathring{D}_{\alpha}\psi_{\mu}-\mathring{D}_{\mu}\psi_{\alpha}\right)+\mathring{D}_{\mu}\left(\xi^{\alpha}\psi_{\alpha}\right)
+\frac{1}{4}\mathring{D}_{[a}\xi_{b]}\gamma^{ab}\psi_{\mu}, \label{Kosmann_RS_1}
\eea
where in the third line we have added and subtracted appropriate terms to bring it in the final form. Later we will convert this final form in the index-free form notation. 

 Jacobson and Mohd \cite{Jacobson:2015uqa} proposed to use the Kosmann derivative to define a Lorentz-diffeo-morphism associated to $\xi$.   Although, the association of the (projection of the) bundle automorphism $\mathcal{K}_\xi $ to spacetime diffeomorphism $\mathcal{L}_\xi$ is different from the spirit of \cite{Prabhu:2015vua}, when restricted to Killing vectors the conclusions from the two formalisms are identical. In particular, one gets the same first law for stationary spacetimes using either of the two prescriptions. We make use of this fact and borrow convenient  notation from both these references to discuss the first law with the Majorana and Rarita-Schwinger fields. 

A comment about notation is in order: Prabhu~\cite{Prabhu:2015vua} uses underline to distinguish between fields on the spacetime (base space) from those on the principal bundle. We do not use that notation. For us \emph{all quantities are on the spacetime.} The main technical point we use from~\cite{Jacobson:2015uqa, Prabhu:2015vua} is the modification of the Lie derivative to 
the Kosmann derivative when evaluating the Noether currents.

Let $\bm{L}$ denote the Lagrangian $d$-form.  Spacetime differential forms are denoted with boldface letters and center dot is used to denote the interior derivative: for a $p$-form $\bm{A}$, $\xi \cdot \bm{A} = i_\xi \bm{A}$. We assume that the Lagrangian is diffeomorphism covariant and is a Lorentz scalar, and its variations are the same when we vary the fields with the Kosmann derivative or with the Lie derivative, i.e.,  
\be
\mathcal{K}_\xi \bm{L} = \mathcal{L}_\xi \bm{L} = \bm{d}(\xi \cdot \bm{ L}).
\ee
Let us denote the dynamical field collectively as ${\bm \varphi}^\alpha$ (and also simply as $\varphi$), where $\alpha$ denotes any internal indices that the field may carry. The variation of the Lagrangian $d$-form induced by a field variation $\delta {\bm \varphi}^\alpha$ can be written as, 
\be
\delta \bm{L} = \bm{\mathcal{E}}_\alpha (\varphi) \wedge \delta \bm{\varphi}^\alpha + d \bm{\theta}(\varphi, \delta \varphi). 
\ee
The quantity $\bm{\mathcal{E}}_\alpha (\varphi)$ defines the equations of motion, $\bm{\mathcal{E}}_\alpha (\varphi) =0$, and the $(d-1)$ form $\bm{\theta}$ is constructed out of the dynamical fields $\varphi$ and their first variations $\delta \varphi$. Now let us consider the variation of the Lagrangian induced by an arbitrary vector field $\xi$, with field variations,
\be
\delta_\xi {\bm \varphi}^\alpha = \mathcal{K}_\xi {\bm \varphi}^\alpha.
\ee
To such a variation we can associate an $(d-1)$-form called the Lorentz-diffeomorphism Noether current,
\be
\bm{J}_\xi = \bm{\theta}(\varphi, \mathcal{K}_\xi  \varphi) - \xi \cdot \bm{L}.  \label{L-D-Noether_current}
\ee
For all vector fields $\xi$, the Noether current is closed on-shell. This implies \cite{Wald1990, Jacobson:2015uqa, Prabhu:2015vua} that on-shell $\bm{J}_\xi$ is an exact form 
$ \bm{J}_\xi = \bm{d Q}_\xi.$ In this work, we are concerned with the construction of the Lorentz-diffeomorphism Noether charge $\bm{Q}_\xi$ for the theories of interest. 

The Lagrangians considered in this paper are of the form, 
\be
\bm{L} = \bm{L}_\rom{grav} + \bm{L}_\rom{matter}, \label{lagrangian}
\ee
where $\bm{L}_\rom{grav}$ depends on the vielbein one-forms (also called frame-field one forms or co-frames) $\bm{e}^a = e_\mu^a dx^\mu$ and the spin-connection one-forms $\bm{\omega}^{ab} = \omega_\mu^{ab} dx^\mu$.  We exclusively work in the first order formalism, i.e., we treat  $\bm{e}^a$ and $\bm{\omega}^{ab}$ as independent fields. 
We write the \emph{equations of motion terms} in the variation with respect to the co-frames $\bm{e}^a$ and the spin-connection $\bm{\omega}^{ab}$ as,
\begin{align}
( \delta_e \bm{L}_\rom{grav})_\rom{eom}& =  \bm{\mathcal{E}}_a \wedge \delta \bm{e}^a,&
( \delta_e \bm{L}_\rom{matter})_\rom{eom}  & =  - \bm{\mathcal{T}}_a \wedge \delta \bm{e}^a,&
&\bm{\mathcal{E}}_a - \bm{\mathcal{T}}_a  = 0,& \\
%
( \delta_\omega \bm{L}_\rom{grav})_\rom{eom}&=  \bm{\mathcal{E}}_{ab} \wedge \delta \bm{\omega}^{ab},&
( \delta_\omega \bm{L}_\rom{matter})_\rom{eom} &=  - \bm{\mathcal{S}}_{ab} \wedge \delta \bm{\omega}^{ab},&
& \bm{\mathcal{E}}_{ab} - \bm{\mathcal{S}}_{ab}  = 0. & 
\end{align}
The above equations define the symbols $\bm{\mathcal{E}}_a, \bm{\mathcal{T}}_a, \bm{\mathcal{E}}_{ab}$ and $\bm{\mathcal{S}}_{ab}$.

For the first order formulation of gravity with a cosmological constant we use the following Lagrangian,
\be
\bm{L}_\rom{grav}=\frac{\left(R-2\Lambda\right)}{16 \pi G} \bm{\epsilon}_{4}~, \label{Lag_grav}
\ee
often called the Einstein-Hilbert-Palatini Lagrangian.  This theory was considered in detail in references~\cite{Jacobson:2015uqa, Prabhu:2015vua}. For this Lagrangian we have a symplectic potential\footnote{Sometimes the $d-1$ form $\bm{\theta}(\varphi, \delta \varphi)$  is called the \emph{presymplectic} potential. Using this we construct the \emph{presymplectic} current $\bm{\omega}(\varphi, \delta_1 \varphi, \delta_2 \varphi) = \delta_1 \bm{\theta}(\varphi, \delta_2 \varphi)-\delta_2 \bm{\theta}(\varphi, \delta_1 \varphi)$. Integrating $\bm{\omega}$ on a Cauchy surface gives a presymplectic 2-form on  the configuration space $\cF$.
The presymplectic 2-form is not a true symplectic 2-form as it is not non-degenerate. It is possible to construct a phase space with a non-degenerate symplectic form~\cite{Lee:1990nz}. However, for our purposes the presymplectic forms will be sufficient. Thus, to simplify terminology we use the names symplectic potential for $\bm{\theta}$ and symplectic current for ${\bm{\omega}}$.} boundary term, which takes the form,
\be
\bm{\theta}  = \frac{1}{32 \pi G} \varepsilon_{abcd} \ \bm{e}^a \wedge \bm{e}^b \wedge \delta \bm{\omega}^{cd}.
\ee

Now, in order to compute the Lorentz-diffeomorphism Noether current \ref{L-D-Noether_current} we need the Kosmann derivative of the spin-connection.  This can be obtained in a number of ways. We follow the following logic: We recall that infinitesimal variation of the spin-connection $\delta \bm{\omega}^{ab}$ transforms covariantly under local Lorentz transformations. The first Cartan structure  equation ($d \bm{e}^{a} +\bm{\omega}^{a}{}_{b} \wedge \bm{e}^{b} = \bm{T}^{a}$, where  $ \bm{T}^{a}$ is the torsion 2-form) allows us to write $\delta \bm{\omega}^{ab}$ in terms of $\delta \bm{e}^{a}$ and $\delta \bm{T}^{a}$. Since  $\bm{T}^{a}$ is a proper Lorentz-tensor, its Kosmann derivative is uniquely defined by the above considerations, cf.~\ref{Kosmann_general}. Therefore, taking $\delta \bm{\omega}^{ab} = \mathcal{K}_\xi \bm{\omega}^{ab}$, $ \mathcal{K}_\xi \bm{\omega}^{ab}$ can be related to $\mathcal{K}_\xi \bm{e}^{a}$ and $\mathcal{K}_\xi \bm{T}^{a}$. A calculation gives, 
\be
\mathcal{K}_\xi \bm{\omega}^{ab} = \mathcal{L}_\xi \bm{\omega}^{ab} - \bm{D} \left(e^{\mu[a} \mathcal{L}_\xi e^{b]}_{\mu} \right),
\ee
where $D$ is the standard local Lorentz covariant derivatives. When restricted to the torsionless case, this equation is same as the one given in \cite{Jacobson:2015uqa}. 

The Lorentz-diffeomorphism Noether current for the gravity sector Lagrangian \ref{Lag_grav} takes the form
\bea
\bm{J}_\xi&=&\frac{1}{32\pi G} \varepsilon_{abcd} \ \bm{e}^a \wedge \bm{e}^b \wedge \left(\mathcal{K}_\xi \bm{\omega}^{cd}\right)-\left(\xi \cdot \bm{L}_\rom{grav}\right)
\\
&=&\bm{d Q}_\xi+ \bm{\mathcal{E}}_{a}\left(\xi \cdot \bm{e}^{a}\right)
+\bm{\mathcal{E}}^{ab}\left\{\frac{1}{2}E_a \cdot E_b \cdot \left[d \bm{\xi}-\bm{T}^c \left(E_c \cdot \bm{\xi}\right)\right]
-\left(E_c \cdot \bm{\xi}\right)\left(E_{[a} \cdot E^c \cdot \bm{T}_{b]}\right)\right\}, \nn
\eea
where
\bea
\bm{\mathcal{E}}_{a} &=& -\frac{1}{16\pi G}\varepsilon_{abcd} \ \bm{e}^b \wedge \bm{R}^{cd}+\frac{\Lambda}{8\pi G}\left(\star \bm{e}_{a}\right)~, \label{starE}
\\
\bm{\mathcal{E}}_{ab} &=& \frac{1}{16\pi G}\varepsilon_{abcd} \ \bm{e}^c \wedge \bm{T}^{d},
\eea
and the Noether charge $\bm{Q}_\xi$ is 
\be
\bm{Q}_\xi= \frac{1}{32\pi G} \varepsilon_{ab}{}^{cd} \ \bm{e}^a \wedge \bm{e}^b \left\{\frac{1}{2}E_c \cdot E_d \cdot \left[d \bm{\xi} - \bm{T}^e \left(E_e \cdot \bm{\xi}\right)\right]-\left(E_e \cdot \bm{\xi}\right)\left(E_{[c} \cdot E^e \cdot \bm{T}_{d]}\right)\right\}.\label{Q_grav_I}
\ee
In equation \ref{starE}, $\star$ denotes the four-dimensional Hodge dual. Expression \ref{Q_grav_I} can be written more conveniently using the contorsion tensor (see below) and can be compared with (the projection of) equation (5.8) of reference \cite{Prabhu:2015vua}. In the torsionless case, the Noether charge $\bm{Q}_\xi$ takes the familiar form, 
\be
\bm{Q}_\xi=-\frac{1}{16\pi G} \left(\star d \bm{\xi}\right).
\ee
To get the first law for a stationary, axisymmetric, black hole solution we use the horizon generating Killing field $k$ in place of $\xi$. Since $k$ vanishes at the bifurcation surface, the integration of the Noether charge  \ref{Q_grav_I} over the bifurcation 2-sphere $\mathcal{B}$ gives \be
\int_\mathcal{B} \bm{Q}_k = \frac{\kappa_B}{8\pi} \ \mathrm{Area}(\mathcal{B}), 
\ee where $\kappa_B$ is surface gravity of the black hole horizon.

\section{Lorentz-Diffeomorphism Noether charge for Majorana field}\label{sec:Majorana}

In this section we study the Lorentz-diffeomorphism Noether charge for the spin-1/2 Majorana field.   The aim of this section is to first compute the symplectic potential $(d-1)$-form $\bm{\theta}$ and then the  corresponding Lorentz-diffeomorphism Noether current $\bm{J}_\xi$.

The Lagrangian four-form for a massive Majorana field takes the form,
\bea
\bm{L}_\rom{matter}  =\bm{\epsilon_4} \ \left(-\frac{1}{2}\bar{\Psi}\slashed{D} \Psi+\frac{1}{2}m \bar{\Psi}\Psi\right)~, \label{Lag_Majorana}
\eea
where $\bar \Psi$ is the Majorana conjugate and the derivative $\slashed{D}$ is defined below. For the benefit of the reader we review the definition of Dirac, Majorana, and charge conjugate following~\cite{Freedman:2012zz}. Let $C$ be the charge conjugation matrix and let $\Gamma^{(r)}$ be the rank $r$  product of antisymmetrised $\gamma$ matrices
\be
\Gamma^{(r)}: \qquad  \gamma^{\mu_1 \ldots \mu_r}  =  \gamma^{[\mu_1} \ldots \gamma^{\mu_r]}. 
\ee 
The conventions reference~\cite{Freedman:2012zz} follows for four-dimensions (and Lorentzian signature) are
\be
(C\Gamma^{(r)})^T = - t_r C\Gamma^{(r)},
\ee
with $t_0 = +1,~t_1 = -1,~t_2 = -1,~t_3 = +1,$ and $t_{r+4} = t_r$. The Majorana conjugate is defined as,
\be
\bar \Psi = \Psi^T C.
\ee
The Majorana field by definition satisfies the Majorana condition, 
$\Psi^C = \Psi$, where $\Psi^C$ is the charge conjugate defined via
$\Psi^* =  i t_0 C \gamma^0 \Psi^C$. These definitions imply that the Dirac conjugate $\bar \Psi = i \Psi^\dagger \gamma^0$ is same as the Majorana conjugate for a Majorana spinor. Due to this equivalence, $\bar{\Psi}$ in \ref{Lag_Majorana} can be taken to be the Dirac conjugate for the Majorana spinor $\Psi$. For explicit computations it is easier to work with the Dirac conjugate. 

The covariant derivative $\slashed{D}$ used in the above Lagrangian is defined as
\be
\slashed{D}\Psi = \gamma^\mu  D_\mu \Psi= \gamma^a E_a^\mu D_\mu \Psi= \gamma^a (E_a \cdot \bm{D} \Psi) =  \gamma^\mu \left( \partial_\mu + \frac{1}{4}\omega_\mu^{~ab} \gamma_{ab}\right) \Psi~,
\ee
where the latin indices are Lorentz frame indices and the greek indices are spacetime indices. 
 It is convenient to define the derivative operator acting on $\bar{\Psi}$ as well via
  \be
 \overline{\slashed{D} \Psi} = - \bar{\Psi} \overleftarrow{\slashed{D}} = - \bar \Psi \overleftarrow{D}_\mu \gamma^\mu =  -\left(\partial_\mu  \bar{\Psi} - \frac{1}{4}\bar{\Psi}\omega_\mu^{~ab} \gamma_{ab}\right) \gamma^\mu~.
 \ee

Unlike the case of the Dirac field, where the reality condition on the Lagrangian requires us to work with the symmetrised derivative $(\bar \Psi \slashed{D} \Psi + \overline{\slashed{D} \Psi}  \Psi)$, Lagrangian \ref{Lag_Majorana} is automatically real due to the Majorana condition. To see this we note that, 
\bea
(\bar \Psi \slashed{D} \Psi)^\dagger &=& (\slashed{D} \Psi)^\dagger (i \Psi^\dagger \gamma^0 )^\dagger  \nn \\ 
&=& i \left(\partial_\mu \Psi ^\dagger + \frac{1}{4} \omega_{\mu ab}  \Psi^\dagger \gamma^0 \gamma^{ab} \gamma^0 \right)(\gamma^0 \gamma^\mu \gamma^0 ) (\gamma^0 \Psi) \nn \\
&=& -\left(\partial_\mu \bar{\Psi} - \frac{1}{4} \omega_{\mu ab} \bar{\Psi} \gamma^{ab}\right) \gamma^\mu \Psi \nn \\
&=& \bar{\Psi} \gamma^\mu \partial_\mu \Psi + \frac{1}{4} \omega_{\mu ab} \bar{\Psi} \gamma^\mu \gamma^{ab} \Psi \nn \\
&=& \bar{\Psi} \slashed{D} \Psi, \label{Maj_real}
\eea
where we have repeatedly used the relation $\bar{\Psi} \gamma^\mu \Psi = 0$ for a Majorana spinor. Note that this property is not available for the Dirac field.

The Lagrangian presented in  \ref{Lag_Majorana} depends on three fields: the vielbein $\bm{e}^{a}$,  the spin-connection $\bm{\omega}^{ab}$ and the Majorana field $\Psi$. The variation of the Lagrangian with respect to the vielbein takes the form,
\bea\label{Maj_EMT}
\delta_{e} \bm{L}&=&(\delta \bm{\epsilon}_{4}) \left(-\frac{1}{2}\bar{\Psi}\slashed{D} \Psi+\frac{1}{2}m \bar{\Psi}\Psi \right) -\frac{1}{2} \bm{\epsilon_{4}} \, \delta (\bar{\Psi} \slashed{D}\Psi)
\nn
\\
&=&(\delta \bm{\epsilon}_{4})\left(-\frac{1}{2}\bar{\Psi}\slashed{D} \Psi+\frac{1}{2}m \bar{\Psi}\Psi\right)
-\frac{1}{2} \bm{\epsilon_{4}}  \, \bar{\Psi}\gamma^{a}(\delta E_{a}\cdot\bm{D}\Psi)
\nn
\\
&=&(\delta \bm{\epsilon}_{4})\left(-\frac{1}{2}\bar{\Psi}\slashed{D} \Psi+\frac{1}{2}m \bar{\Psi}\Psi\right)+\frac{1}{2} \bm{\epsilon_{4}} \  \bar{\Psi}\gamma^{a} (E_{b}\cdot\bm{{D}}\Psi)(E_{a}\cdot \delta \bm{e}^{b})
\nn
\\
&=&-\left(\star\bm{e}_{a}\right)\left(-\frac{1}{2}\bar{\Psi}\slashed{D} \Psi+\frac{1}{2}m \bar{\Psi}\Psi \right) \w \delta \bm{e}^{a} -\frac{1}{2} (E_{b} \cdot \bm{\epsilon_{4}}) {\bar{\Psi}}\gamma^{b}( E_{a}\cdot{\bm{D}}\Psi)\w \delta \bm{e}^{a}
\nn
\\
&=&-\left(\star\bm{e}_{a}\right)\left(-\frac{1}{2}\bar{\Psi}\slashed{D} \Psi+\frac{1}{2}m \bar{\Psi}\Psi\right)\w \delta \bm{e}^{a} -\frac{1}{2} \star\bm{e}_{b} \ \bar{\Psi} \gamma^{b}( E_{a} \cdot \bm{D}\Psi) \w \delta \bm{e}^{a}
\nn
\\
&:=& -\bm{\Tau}_{a}\w \delta \bm{e}^{a}~.
\eea
The above manipulations have been performed as follows. In going from the second to the third step we have related $\delta E_{a}$ to $\delta \bm{e}^{b}$. In going from the third to the  fourth step we have used the result, $(\delta \bm{\epsilon}_{4}) = -(\star\bm{e}_{a}) \w \delta \bm{e}^{a}$, where $\star$ denotes the Hodge star, and also the identity,
\be
V \cdot (\bm{A}\wedge \bm{B})=(V \cdot \bm{A})\wedge \bm{B}+ (-1)^p \bm{A} \wedge (V \cdot\bm{B})~, 
\label{identity}
\ee
where, $V$ is a vector field, $\bm{A}$ is a $p$-form and $\bm{B}$ is $q$-form. In the present context we have used identity \ref{identity} with $ \bm{A} =  \bm{\epsilon}_{4}$, $\bm{B} = \delta \bm{e}^{a}$ and $V= E_{a}$, such that, $\bm{A}\wedge \bm{B}=0$. In going from the fourth to the fifth step we have used $(E_{b} \cdot \bm{\epsilon_{4}})=\star\bm{e}_{b}$. Finally, the last line of \ref{Maj_EMT} defines the energy-momentum three-form $\bm{\Tau}_{a}$, 
\bea
\bm{\Tau}_{a}=\left(\star\bm{e}_{a}\right)\left(-\frac{1}{2} \bar{\Psi} \slashed{D} \Psi+\frac{1}{2} m \bar{\Psi} \Psi \right)+\frac{1}{2} \left(\star\bm{e}_{b}\right)\bar{\Psi} \gamma^{b} E_{a} \cdot \bm{D} \Psi~. \label{EOM_a}
\eea

Similarly, the variation with respect to the spin connection gives,
\bea
\delta_{\omega} \bm{L}&=&-\frac{1}{2} \bm{\epsilon_{4}} \ \bar{\Psi} \gamma^{a} E_{a}\cdot \left(\frac{1}{4} \delta \bm{\omega}_{cd}\gamma^{cd}\right) \Psi
\nn
\\
&=&-\frac{1}{8} \bm{\epsilon}_{4} \ \bar{\Psi} \gamma^{a} \gamma^{cd} \Psi (E_{a}\cdot \delta \bm{\omega}_{cd})
=\frac{1}{8} (E_{a} \cdot \bm{\epsilon}_{4}) \bar{\Psi} \gamma^{a} \gamma^{cd} \Psi \w \delta \bm{\omega}_{cd} 
\nn
\\
&:=& - \bm{S}^{cd}\w \delta \bm{\omega}_{cd}~,
\eea
where we have again used identity \ref{identity} with $\bm{A}=\bm{\epsilon}_{4}$, $\bm{B} = \delta \bm{\omega}_{cd}$, and $V= E_a$ such that $\bm{A}\wedge \bm{B}=0$. The last line of the above equation defines the spin-current three-form $\bm{S}^{cd}$,
\bea
\bm{S}^{cd}=-\frac{1}{8}(E_{a} \cdot \bm{\epsilon}_{4}) \ \bar{\Psi} \gamma^{a} 
\gamma^{cd} \Psi. \label{EOM_ab}
\eea

Finally, we  consider the variation of the Lagrangian with respect to $\Psi$. Unlike the case of Dirac field, where variations of $\Psi$ and $\bar{\Psi}$ are treated as independent, here the field variations must also satisfy the Majorana condition so that $\delta \Psi$ is related to $\delta \bar \Psi$. In practice this condition is implemented through the \emph{Majorana flip} relations. Let $\lambda_1$ and $\lambda_2$ be two arbitrary Majorana spinors (possibly with other spacetime indices), then, 
\be
\bar{\lambda}_1 \, \Gamma^{(r)} \lambda_2 = t_r \, \bar{\lambda}_2 \, \Gamma^{(r)} \lambda_1, \label{majorana_flip}
\ee
which for our applications read,
\begin{align}
& \delta \bar{ \Psi} \Psi = \bar{ \Psi} \delta \Psi~, & 
&\delta \bar{ \Psi}  \gamma_{a} \Psi = - \bar{ \Psi} \gamma_{a} \delta \Psi~, \\
&\delta \bar{ \Psi}\gamma_{cd} \Psi = - \bar{ \Psi} \gamma_{cd} \delta \Psi~, &
&\delta \bar{ \Psi}\gamma_{abc} \Psi = \bar{ \Psi} \gamma_{abc} \delta \Psi~, \\
&\delta \bar{ \Psi} \gamma^a \partial_\mu  \Psi = -  \partial_\mu \bar{ \Psi} \gamma^a  \delta \Psi~, &
& \delta \bar{ \Psi}\gamma_{a} \gamma_{cd} \Psi=\bar{ \Psi} \gamma_{cd} \gamma_{a}  \delta \Psi.
\end{align}
These relations imply,
\bea
\delta \bar{\Psi} \slashed{D} \Psi &=& \delta \bar{\Psi} \gamma ^{\mu}\left( \partial_\mu + \frac{1}{4}\omega_\mu^{~ab} \gamma_{ab}\right) \Psi
\nn
\\
&=&-\partial_\mu\bar{\Psi} \gamma ^{\mu} \delta \Psi+ \frac{1}{4}\bar{\Psi}\omega_\mu{}^{ab} \gamma_{ab}\gamma^{\mu}\delta \Psi \nn \\
&=&-\bar{\Psi} \overleftarrow{\bm{D}} \cdot E_a \gamma^a \delta \Psi =-\bar{\Psi} \overleftarrow{\slashed{D}} \delta \Psi ~. \label{left_D_2}
\eea

Using these relations, the variation of the Lagrangian with respect to the Majorana field yields,  
\bea\label{Maj_Boundary}
\delta_{\Psi}\bm{L}&=&\frac{1}{2}\bm{\epsilon_{4}}\left(-\delta \bar{\Psi} \slashed{D} \Psi - \bar{\Psi}\slashed{D}\delta \Psi + m {\bar{\Psi}}\delta \Psi + m \delta{\bar{\Psi}} \Psi \right)
\nn
\\
&=& \frac{1}{2} \bm{\epsilon_{4}} \bar{\Psi} \overleftarrow{\slashed{D}} \delta \Psi +  \frac{1}{2}(E_{a}\cdot \bm{\epsilon_{4}}){\bar{\Psi}} \gamma^{a} \w \bm{D}\delta  \Psi+\bm{\epsilon_{4}}(m \bar{\Psi}\delta \Psi) 
\nn
\\ 
&=& \bm{\epsilon_{4}} \left( \bar{\Psi} (\overleftarrow{\slashed{D}} + m) -\frac{1}{3!} T_{ba}{}^b (\bar{\Psi} \gamma^a)  \right) \delta \Psi - \bm{d}\left[\frac{1}{2}(E_{a}\cdot \bm{\epsilon_{4}})\bar{\Psi} \gamma^{a} \delta \Psi \right],
\eea
where $T_{ab}{}^{c}$ are the frame components of the torsion $T_{ab}{}^{c} = E_b \cdot E_a \cdot \bm{T^{c}}$.  In going from the first to the second step we have used once again identity \ref{identity} with $\bm{A}=\bm{\epsilon}_{4}$, $\bm{B}=\bm{D}\delta \Psi$, and $V= E_a$ such that $\bm{A}\wedge \bm{B}=0$. In going from the second to the third line we have used integration by parts and manipulations similar to the ones performed in \cite{Prabhu:2015vua} for the Dirac field.

The contribution to the symplectic potential $\bm{\theta}$ from the Majorana field can be read off from \ref{Maj_Boundary},
\bea
\bm{\theta}(\Psi,\delta\Psi)= -\frac{1}{2}(E_{a}\cdot \bm{\epsilon_{4}})\bar{\Psi} \gamma^{a} \delta \Psi~.
\eea
Given the symplectic potential, the Lorentz-diffeomorphism Noether current can be obtained in a straight forward manner. It takes the form,
\bea
\bm{J}_\xi&:=& \bm{\theta}(\Psi,\mathcal{K}_{\xi}{\Psi})-\xi\cdot \bm{L}
\nn
\\
&=&-\frac{1}{2}(E_{a}\cdot \bm{\epsilon_{4}})\bar{\Psi}\gamma^{a} \left( \mathcal{K}_{\xi}\Psi \right) -(\xi \cdot \bm{\epsilon}_{4})\left(-\frac{1}{2}\bar{\Psi}\slashed{D} \Psi+\frac{1}{2}m \bar{\Psi}\Psi\right). \label{J_intermediate}
\eea
From \ref{Kosmann_spinor} we have, 
\be
\mathcal{K}_{\xi}\Psi = \xi \cdot \bm{d} \Psi + \frac{1}{4} \xi \cdot \mathring{\bm{\omega}}_{ab} \gamma^{ab} \Psi + \frac{1}{8} \left( E_b \cdot E_a \cdot d \bm{\xi} \right) \gamma^{ab}\Psi,
\ee
Inserting this expression in \ref{J_intermediate} and using the identity, 
\be
V \cdot {\bm A} = (V \cdot \bm{e}_a )(E^a \cdot {\bm A}),
\ee
we get
\be
\bm{J}_\xi =  -\bm{\Tau}_{a} (\xi \cdot \bm{e}^{a}) - \bm{\mathcal{S}}^{ab} \left(\frac{1}{2}E_a \cdot E_b \cdot \left(d \bm{\xi} - \bm{T}^c (E_c \cdot \bm{\xi}) \right) - (E_c \cdot \bm{\xi}) (E_{[a} \cdot E^c \cdot \bm{T}_{b]}\right),
\ee
where $\bm{\Tau}_{a}$ and $\bm{\mathcal{S}}_{ab}$ are defined in \ref{EOM_a} and  \ref{EOM_ab} respectively.

Now, the full Lorentz-diffeomorphism Noether current of the gravity plus the matter takes the form, 
\bea
\bm{J}_\xi &=& \bm{J}^{\rom{grav}}_\xi + \bm{J}^{\rom{matter}}_\xi \\
&=&  \bm{d Q}_\xi +  \left( \bm{\mathcal{E}}_{a} -\bm{\mathcal{T}}_{a} \right)  (\xi \cdot \bm{e}^{a}) \nn \\ 
 && +\left( \bm{\mathcal{E}}^{ab} -\bm{\mathcal{S}}^{ab} \right)  \left(\frac{1}{2}E_a \cdot E_b \cdot \left(d \bm{\xi} - \bm{T}^c (E_c \cdot \bm{\xi}) \right) - (E_c \cdot \bm{\xi}) (E_{[a} \cdot E^c \cdot \bm{T}_{b]}\right).
\eea
Therefore, on-shell we have $\bm{J}_\xi= \bm{d Q}_\xi $ where  the Noether charge $\bm{Q}_\xi$ is 
\be
\bm{Q}_\xi = \frac{1}{32 \pi G} \varepsilon_{ab}{}^{cd} \ \bm{e}^a \wedge \bm{e}^b \left(\frac{1}{2}E_c \cdot E_d \cdot \left(d \bm{\xi} - \bm{T}^e (E_e \cdot \bm{\xi}) \right) - (E_e \cdot \bm{\xi}) (E_{[c} \cdot E^e \cdot \bm{T}_{d]}\right),
\ee
and the torsion 2-form is fixed by the equations of motion $ \bm{\mathcal{E}}^{ab} -\bm{\mathcal{S}}^{ab} =0.$ At the bifurcation surface, for the horizon generating Killing field  $\xi^\mu = k^\mu =  (\partial_t)^\mu  + \Omega (\partial_\phi)^\mu $,  $\xi^\mu $ vanishes. As a result, the Noether charge $\bm{Q}_\xi$ simplifies to 
\be
\bm{Q}_k \bigg{|}_\mathcal{B} = \frac{1}{32 \pi G} \varepsilon_{ab}{}^{cd} \ \bm{e}^a \wedge \bm{e}^b \left( \frac{1}{2}E_c \cdot E_d \cdot d \bm{k} \right) = - \frac{1}{16 \pi G} \star d \bm{k}.
\ee
Thus, we see that the Majorana fermion does not contribute to the full Noether charge for the horizon generating Killing field at the bifurcation surface. 

Note that, through the change in  torsion due to the Majorana field,  in general both the Noether current and the Noether charge  are different from pure general relativity. However, when the Noether charge is evaluated for the horizon generating Killing field at the bifurcation surface all those terms do not contribute.

\section{Lorentz-Diffeomorphism Noether charge for simple (AdS) supergravity}\label{sec:supergravity}

In this section we study the Lorentz-diffeomorphism Noether charge for the spin-3/2 Rarita-Schwinger field in four spacetime dimensions. The most natural set-up to consider is the case of $N=1, D=4$ supergravity, often called simple supergravity.  
 With very little effort, this discussion can be generalised to $N=1, D=4$ AdS supergravity. This is the set-up we work with. The Lagrangian 4-form for simple AdS supergravity is,
 \bea
 {\bm L}_\rom{sugra} = \frac{1}{4 \kappa^2} \varepsilon_{abcd} \ {\bm R}^{ab}(\bm{\omega}) \wedge \bm{e}^c \wedge \bm{e}^d + \frac{i}{2 \kappa^2} \bar {\bm{\psi}}\w \gamma_{*} \bm{\gamma}\w \bm{\hat{D}}\bm{\psi} -  \frac{1}{\kappa^2} \Lambda \, \bm{\epsilon_4}, \label{lagrangian_sugra}
 \eea
 where the Rarita-Schwinger field $\bm{\psi}$ is a Majorana spinor valued one-form, $\bm{\gamma}$ is a matrix-valued one-form 
  \be
\bm{\gamma} = \gamma^a \bm{e}_a,
 \ee
 $\gamma_{*}=i\gamma_{0}\gamma_{1}\gamma_{2}\gamma_{3}$, and the derivative operator on the Rarita-Schwinger field is defined as,  
  \be
\bm{\hat{D}}\bm{\psi} = \bm{D}\bm{\psi}- \frac{1}{2L} \bm{\gamma} \wedge \bm{\psi} = \left(\partial_{\mu}\psi_{\nu}+\frac{1}{4}\omega_{\mu}^{~cd}\gamma_{cd}\psi_{\nu} - \frac{1}{2L} \gamma_\mu \psi_\nu \right)\bm{d}x^{\mu}\w \bm{d}x^{\nu}.
 \ee
 Furthermore, $\kappa^2 = 8 \pi G$ and the cosmological constant $\Lambda$ is related to the AdS radius $L$ through $\Lambda = -\frac{3}{L^2}$.   Lagrangian \ref{lagrangian_sugra} is presented in a slightly different form compared to standard references such as \cite{Freedman:2012zz}. Since we use the form notation, the above structure of the Lagrangian is computationally simpler to work with.

The gravity sector of the Lagrangian presented in \ref{lagrangian_sugra} is the same as in the previous sections. Therefore, it is enough to consider the matter part of the Lagrangian \ref{lagrangian_sugra},
\be
 {\bm L} =   \frac{i}{2 \kappa^2} \bar {\bm{\psi}}\w \gamma_{*} \bm{\gamma}\w \bm{\hat{D}}\bm{\psi}.
\ee

The variation with respect to $\bm{e}^{a}$  yields the energy-momentum 3-form $\bm{\mathcal{T}}_a$, and the variation with respect to $\bm{\omega}^{ab}$  yields the spin-current 3-form $\bm{\mathcal{S}}_{ab}$,
\begin{align}
( \delta_e \bm{L})_\rom{eom} & =  - \bm{\mathcal{T}}_a \wedge \delta \bm{e}^a,&
(\delta_\omega \bm{L})_\rom{eom} &=  - \bm{\mathcal{S}}_{ab} \wedge \delta \bm{\omega}^{ab}.
\end{align}
It turns out that the variations $ \delta_e \bm{L}$ and $\delta_\omega \bm{L}$ do not give total derivative terms.
The variation with respect to $\bm{\psi}$  yields the equations of motion for the Rarita-Schwinger field and a  total derivative term from where we read off the contribution to the symplectic potential 3-form $\bm{\theta}$.

We have,
\bea
\delta \bm{L}&=&+\frac{i}{2 \kappa^2}\left(\delta\bar {\bm{\psi}}\right)\w \gamma_{*} \bm{\gamma} \w \bm{\hat{D}}\bm{\psi}
+\frac{i}{2 \kappa^2}\bar {\bm{\psi}}\w \gamma_{*} \gamma_{a} \bm{\hat{D}}\bm{\psi} \w \delta \bm{e}^{a}
\nonumber
\\
&&+\frac{i}{2 \kappa^2}\bar {\bm{\psi}}\w \gamma_{*} \bm{\gamma} \w \bm{\hat{D}}\delta\bm{\psi}
+\frac{i}{2 \kappa^2}\bar {\bm{\psi}}\w \gamma_{*} \bm{\gamma} \w \left(\frac{1}{4}\delta \bm{\omega}^{cd}\gamma_{cd}\right)\w \bm{\psi} \nonumber \\
&& + \frac{i}{2 \kappa^2} \bar {\bm{\psi}}\w \gamma_{*} \bm{\gamma} \w \left(\frac{1}{2 L }\gamma_{a}\right) \bm{\psi}  \w \delta \bm{e}^{a}. \label{RS_var}
\eea
The variation of $\bar {\bm{\psi}}$ is related to the variation of $\bm{\psi}$ through the Majorana flip relations. The following more general version of the Majorana flip relation is  very useful in actual computations: 
\be
\bar{\lambda}_1 \, \Gamma^{(r_1)} \Gamma^{(r_2)} \lambda_2 = t_{r_1} \overline{ \left(\Gamma^{(r_2)} \lambda_2  \right)} \Gamma^{(r_1)} \ \lambda_1= t_0 t_{r_1} t_{r_2} \, \bar{\lambda}_2 \, \Gamma^{(r_2)} \Gamma^{(r_1)} \lambda_1. \label{majorana_flip2}
\ee
Using $\gamma_{*}=i\gamma_{0}\gamma_{1}\gamma_{2}\gamma_{3}$ together with \ref{majorana_flip} and  \ref{majorana_flip2}, 
  we have for the first term in the variation presented in  \ref{RS_var},
\bea
\left(\delta\bar {\bm{\psi}}\right)\w \gamma_{*}\gamma_{a}\bm{e}^{a}\w \hat{\bm{D}}\bm{\psi}
&=&\left(\delta \bar{\psi}_{\mu}\gamma_{*}\gamma_{a} \hat{D}_{\rho}\psi_{\sigma} \ e^{a}{}_{\nu} \right)\bm{d}x^{\mu}\w \bm{d}x^{\nu}\w \bm{d}x^{\rho}\w \bm{d}x^{\sigma}
\nonumber
\\
&=&\left( \overline{(\hat{D}_{\rho}\psi_{\sigma})} \ \gamma_{*}\gamma_{a} \delta \bar{\psi}_{\mu} \ e^{a}{}_{\nu} \right)\bm{d}x^{\mu}\w \bm{d}x^{\nu}\w \bm{d}x^{\rho}\w \bm{d}x^{\sigma}
\nonumber
\\
&=&\left\{ \left(\partial_{\rho}\bar{\psi}_{\sigma} - \frac{1}{4} \omega_{\rho cd} \bar{\psi}_{\sigma} \gamma^{cd} + \frac{1}{2 L} \bar{\psi}_{\sigma} \gamma_{\rho} \right) \ \gamma_{*}\gamma_{\nu} \ \delta \psi_{\mu} \right\} \bm{d}x^{\sigma}\w \bm{d}x^{\rho}\w \bm{d}x^{\nu}\w \bm{d}x^{\mu}
\nonumber
\\
&=& - \left( \bm{d} \bar{\bm{\psi}} + \frac{1}{4} \bar{\bm{\psi}} \w \bm{\omega}_{cd} \gamma^{cd} - \frac{1}{2L} \bar{\bm{\psi}} \w \bm{\gamma} \right) \w \gamma_{*} \bm{\gamma} \w \delta \bm{\psi}
\nonumber
\\
&=&-\bar{\bm{\psi}}\overleftarrow{\hat{\bm{D}}}\w \gamma_{*}\gamma_{a}\bm{e}^{a}\w \delta \bm{\psi}, \label{RS_flip}
\eea 
where the operation $\overleftarrow{\hat{\bm{D}}}$ is defined as,
\bea
\bar{\bm{\psi}}\overleftarrow{\hat{\bm{D}}} &\equiv& \bm{d} \bar{\bm{\psi}} + \frac{1}{4} \bar{\bm{\psi}} \w \omega_{cd} \gamma^{cd} - \frac{1}{2 L} \bar{\bm{\psi}} \w \bm{\gamma} 
\nonumber 
\\
&=& \left(\partial_{\mu}\bar{\psi}_{\nu}-\frac{1}{4}\omega_{\mu cd}\bar{\psi}_{\nu}\gamma^{cd} + \frac{1}{2 L} \bar{\psi}_{\nu} \gamma_{\mu} \right)\bm{d}x^{\mu}\w \bm{d}x^{\nu} 
\nonumber 
\\
&=& \bar{\psi}_{\nu}\overleftarrow{\hat{D}_{\mu}} \bm{d}x^{\mu}\w \bm{d}x^{\nu}.
\eea
Additionally, to simplify the second term in the variation \ref{RS_var}, we note that 
\bea
\bar{\bm{\psi}} \w \gamma_{*} \bm{\gamma} \w \hat{\bm{D}} \delta \bm{\psi} &=& \bar{\bm{\psi}} \w \gamma_{*} \bm{\gamma} \w \bm{d} \delta \bm{\psi} + \frac{1}{4} \bar{\bm{\psi}} \w \gamma_{*} \bm{\gamma} \w \bm{\omega}_{cd} \gamma^{cd} \w \delta \bm{\psi} - \frac{1}{2 L} \bar{\bm{\psi}} \w \gamma_{*} \bm{\gamma} \w \bm{\gamma} \w \delta \bm{\psi} 
\nonumber
\\
&=& \bm{d} \left( \bar{\bm{\psi}} \w \gamma_{*} \bm{\gamma} \w \delta \bm{\psi} \right) - \bm{d} \left( \bar{\bm{\psi}} \w \gamma_{*} \bm{\gamma} \right) \w \delta \bm{\psi} 
\nonumber 
\\
&& - \frac{1}{4} \bar{\bm{\psi}} \w \bm{\omega}_{cd} \w \gamma_{*} \bm{\gamma} \gamma^{cd} \w \delta \bm{\psi} + \frac{1}{2 L} \bar{\bm{\psi}} \w \bm{\gamma} \w \gamma_{*} \bm{\gamma} \w \delta \bm{\psi}
 \nonumber
\\
&=& \bm{d} \left( \bar{\bm{\psi}} \w \gamma_{*} \bm{\gamma} \w \delta \bm{\psi} \right) +    \bar{\bm{\psi}} \w \gamma_{*} \gamma_{a} \bm{d} \bm{e}^a  \w \delta \bm{\psi}   
\nonumber
\\
&& - \left( \bm{d}  \bar{\bm{\psi}} - \frac{1}{2 L} \bar{\bm{\psi}} \w \bm{\gamma} \right) \w \gamma_{*} \bm{\gamma} \w \delta \bm{\psi} - \frac{1}{4} \bar{\bm{\psi}} \w \bm{\omega}_{cd} \w \gamma_{*} \gamma^{a} \gamma^{cd} \bm{e}_a \w \delta \bm{\psi}
\nonumber
\\
&=& \bm{d} \left( \bar{\bm{\psi}} \w \gamma_{*} \bm{\gamma} \w \delta \bm{\psi} \right) +    \bar{\bm{\psi}} \w \gamma_{*} \gamma_{a} \left( \bm{d} \bm{e}^a + \bm{\omega}^{a}{}_{b} \w \bm{e}^b \right) \w \delta \bm{\psi} 
\nonumber
\\
&& - \left(\bm{d} \bar{\bm{\psi}} + \frac{1}{4} \bar{\bm{\psi}} \w \bm{\omega}_{cd} \gamma^{cd} - \frac{1}{2 L} \bar{\bm{\psi}} \w \bm{\gamma} \right) \w \gamma_{*} \bm{\gamma} \w \delta \psi
\nonumber
\\
&=& \bm{d} \left( \bar{\bm{\psi}} \w \gamma_{*} \bm{\gamma} \w \delta \bm{\psi} \right) + \bar{\bm{\psi}} \w \gamma_{*} \gamma_{a} \bm{T}^a \w \delta \bm{\psi} - \bar{\bm{\psi}} \overleftarrow{\hat{\bm{D}}} \w \gamma_{*} \bm{\gamma} \w \delta \bm{\psi}, \label{RS_psi_var}
\eea
where we have separated out the total derivative term.

Using both \ref{RS_flip} and \ref{RS_psi_var}, the variation of the Rarita-Schwinger Lagrangian simplifies to,
\bea
\delta \bm{L}&=& 
\bm{d} \bm{\theta} -\bm{\mathcal{T}}_{a}\w \delta\bm{e}^{a}-\bm{\cS}_{cd}\w \delta\bm{\omega}^{cd} -\bm{\cE}_{\bm{\psi}}\w \delta \bm{\psi},
\eea 
with
\begin{align}
&\bm{\theta}=\frac{i}{2\kappa^2}\bar {\bm{\psi}}\w \gamma_{*} \bm{\gamma} \w \delta\bm{\psi}~, \label{new_theta} \\
&\bm{\Tau}_{a} = -\frac{i}{2\kappa^2}\left[\bar {\bm{\psi}}\w \gamma_{*}\gamma_{a} \bm{D}\bm{\psi} -\frac{1}{L}\bar{\bm{\psi}}\w \gamma_{*} \gamma_{ab} \bm{e}^{b} \w \bm{\psi} \right], \label{new_tau}
\\
&\bm{\cS}_{cd}=\frac{i}{8\kappa^2}\bar {\bm{\psi}}\w \gamma_{*} \bm{\gamma} \gamma_{cd} \w \bm{\psi}~, \\
& \bm{\cE}_{\bm{\psi}}=\frac{i}{\kappa^2} \left[ \bar{\bm{\psi}}\overleftarrow{\bm{\hat{D}}}\w \gamma_{*} \bm{\gamma} - \frac{1}{2} \bar{\bm{\psi}} \w \gamma_{*} \gamma_{a} \bm{T}^a \right]. \label{new_eom}
\end{align}

Having obtained an expression for $\bm{\theta}$, we can now obtain the  Lorentz-diffeomorphism Noether current via our general formula,
\be
\bm{J}_\xi=\bm{\theta}(\bm{\psi},\cK_{\xi}\bm{\psi})-\xi\cdot \bm{L}.
\ee
To compute this  we need the Kosmann derivative $\cK_{\xi}\bm{\psi}$ of the Rarita-Schwinger field. An expression for  $\cK_{\xi}\bm{\psi}$ was given in \ref{Kosmann_RS_1}.  Expression \ref{Kosmann_RS_1} can be written more concisely in the form notation as follows,
\be
\mathcal{K}_{\xi} \bm{\psi}=\xi \cdot \mathring{\bm{D}}\bm{\psi}+\mathring{\bm{D}}\left(\xi \cdot \bm{\psi}\right)+\frac{1}{4}\left(-\frac{1}{2}E_{a}\cdot E_{b}\cdot \bm{d\xi} \right) \gamma^{ab}\bm{\psi},
\ee
or equivalently,
\bea
\mathcal{K}_{\xi} \bm{\psi}&=&\xi \cdot \left(\bm{D}\bm{\psi}-\frac{1}{4}\bm{K}_{ab}\gamma^{ab}\w \bm{\psi}\right)
+\bm{D}\left(\xi \cdot \bm{\psi}\right)-\frac{1}{4}\bm{K}_{ab}\gamma^{ab}\left(\xi \cdot \bm{\psi}\right)+\frac{1}{4}\left( -\frac{1}{2}E_{a}\cdot E_{b}\cdot \bm{d\xi} \right) \gamma^{ab}\bm{\psi}
\nonumber
\\
&=&\xi \cdot \left(\bm{D}\bm{\psi}\right)+\bm{D}\left(\xi \cdot \bm{\psi}\right)+\frac{1}{4}\left(-\frac{1}{2}E_{a}\cdot E_{b}\cdot \bm{d\xi} -\xi\cdot \bm{K}_{ab}\right)\gamma^{ab}\bm{\psi}, \label{Kosmann_RS}
\eea 
where we have replaced $\mathring{\bm{D}}$ with $\bm{D}$. In the new form the spinor covariant derivative includes torsion.  The contorsion tensor $\bm{K}_{ab}$ is defined via $\bm{\omega}_{ab} = \mathring{\bm{\omega}}_{ab} + \bm{K}_{ab}$, and can be expressed in terms of the torsion tensor as
\be
K_{\alpha \mu \nu}=-\frac{1}{2}\left(T_{\alpha \mu \nu}-T_{\mu \nu \alpha}+T_{\nu \alpha \mu}\right).
\ee
The contorsion tensor is antisymmetric in the last two indices, while torsion tensor is antisymmetric in the first two indices.  We can write contorsion tensor term $\xi \cdot \bm{K}_{ab}$  in terms of  the torsion two form. We have, 
\bea
\xi \cdot \bm{K}_{ab}&=&\xi^{\alpha}K_{\alpha \mu \nu}E^{\mu}_{a}E^{\nu}_{b}
=-\frac{1}{2}\xi^{\alpha}\left(T_{\alpha \mu \nu}-T_{\mu \nu \alpha}+T_{\nu \alpha \mu}\right)E^{\mu}_{a}E^{\nu}_{b}
\nonumber
\\
&=&\frac{1}{2}\xi^{\alpha}T_{\mu \nu \alpha}E^{\mu}_{a}E^{\nu}_{b}-\frac{1}{2}\xi^{\alpha}\left(T_{\alpha \mu \nu}-T_{\alpha \nu \mu}\right)E^{\mu}_{a}E^{\nu}_{b}
\nonumber
\\
&=&-\frac{1}{2}\left(\xi \cdot \bm{e}^{c}\right)\left[E_{a}\cdot \left(E_{b}\cdot \bm{T}_{c}\right)\right]-E_{[a}\cdot \left(\xi \cdot \bm{T}_{b]}\right) \label{K_to_T}
\eea
where $\bm{T}_{a}=(1/2)T_{\mu \nu \alpha}E^{\alpha}_{a}~\bm{d}x^{\mu}\w \bm{d}x^{\nu}$. Thus, the Kosmann derivative of the spinor one-form can be written as: 
\be 
\mathcal{K}_{\xi} \bm{\psi}=\xi \cdot \left(\bm{D}\bm{\psi}\right)+\bm{D}\left(\xi \cdot \bm{\psi}\right)+\frac{1}{4}\left[-\frac{1}{2}E_{a}\cdot \left(E_{b}\cdot \bm{d\xi}\right)
+\frac{1}{2}\left(\xi \cdot \bm{e}^{c}\right)\left[E_{a}\cdot \left(E_{b}\cdot \bm{T}_{c}\right)\right]+E_{[a}\cdot \left(\xi \cdot \bm{T}_{b]}\right)\right]\gamma^{ab}\bm{\psi}. \label{Kosmann_RS_3}
\ee
It is slightly easier to work with expression \ref{Kosmann_RS}, as it is less cumbersome. However, to compare with some expressions in \cite{Prabhu:2015vua} it is better to use the form \ref{Kosmann_RS_3}. We continue to use expression \ref{Kosmann_RS} and use \ref{K_to_T} to convert contorsion tensor into the torsion tensor when needed.

Now, from equations \ref{new_theta} and \ref{new_eom}, it follows that
\bea
\bm{\theta}(\bm{\psi}, \cK_\xi \bm{\psi} ) 
&=& \frac{i}{2 \kappa^2}  \bar{\bm{\psi}} \w \gamma_{*} \bm{\gamma} \w \xi \cdot \bm{D \psi}   + \frac{i}{2 \kappa^2} \bar{\bm{\psi}} \w \gamma_{*} \bm{\gamma} \w \bm{D} (\xi \cdot \bm{\psi}) \nonumber
\\ && - \cS^{cd} \left( \half E_c \cdot E_d \cdot \bm{d \xi} + \xi \cdot \bm{K}_{cd} \right).
\label{partI}
\eea
The $\xi\cdot \bm{L}$ term in  $\bm{J}_\xi=\bm{\theta}(\bm{\psi},\cK_{\xi}\bm{\psi})-\xi\cdot \bm{L}$ is more tedious. We find,
\bea
\xi \cdot \bm{L} &=& \frac{i}{2 \kappa^2} (\xi \cdot \bar{\bm{\psi}}) \gamma_{*} \bm{\gamma} \w \hat{\bm{D}} \bm{\psi} - \frac{i}{2 \kappa^2} \bar{\bm{\psi}} \w \gamma_{*} \gamma_a \hat{\bm{D}} \bm{\psi} (\xi \cdot \bm{e}^a ) \nonumber \\
&& + \frac{i}{2 \kappa^2} \bar{\bm{\psi}} \w \gamma_{*} \bm{\gamma} \w \xi \cdot \bm{D} \bm{\psi} - \frac{i}{2 \kappa^2} \left(\frac{1}{2 L} \right) \bar{\bm{\psi}} \w \gamma_{*} \bm{\gamma} \gamma_a \w \bm{\psi} (\xi \cdot \bm{e}^a) 
\nonumber \\
&& + \frac{i}{2 \kappa^2} \left( \frac{1}{2 L} \right)\bar{\bm{\psi}} \w \gamma_{*} \bm{\gamma} \w \bm{\gamma} (\xi \cdot \bm{\psi}) 
\nn
\\
&=& \frac{i}{2 \kappa^2} (\xi \cdot \bar{\bm{\psi}}) \gamma_{*} \bm{\gamma} \w \hat{\bm{D}} \bm{\psi} + \frac{i}{2 \kappa^2} \bar{\bm{\psi}} \w \gamma_{*} \bm{\gamma} \w \xi \cdot \bm{D} \bm{\psi} 
\nn
\\
&& \left( - \frac{i}{2 \kappa^2} \xi \cdot \bm{e}^a \right) \left[ \bar{\bm{\psi}} \w \gamma_{*} \gamma_{a} \bm{D \psi} - \frac{1}{2 L} \bar{\bm{\psi}} \w \gamma_{*} \left( \gamma_{a} \gamma_{b} - \gamma_{b} \gamma_{a} \right) \bm{e}^b \w \bm{\psi} \right]
\nn
\\
&& + \frac{i}{2 \kappa^2} \left(\frac{1}{2 L} \right) \bar{\bm{\psi}} \w \gamma_{*} \bm{\gamma} \w \bm{\gamma} \ \left(\xi \cdot \bm{\psi} \right)
\nn
\\
&=& \frac{i}{2 \kappa^2} (\xi \cdot \bar{\bm{\psi}}) \gamma_{*} \bm{\gamma} \w \hat{\bm{D}} \bm{\psi} + \frac{i}{2 \kappa^2} \bar{\bm{\psi}} \w \gamma_{*} \bm{\gamma} \w \xi \cdot \bm{D} \bm{\psi} 
\nn
\\
&& + \left( \xi \cdot \bm{e}^a \right) \bm{\cT}_a + \frac{i}{2 \kappa^2} \left(\frac{1}{2 L} \right) \bar{\bm{\psi}} \w \gamma_{*} \bm{\gamma} \w \bm{\gamma} \ \left(\xi \cdot \bm{\psi} \right).
\label{partII}
\eea
The above manipulations are as follows. In the first step we have expanded $\xi \cdot \hat{\bm{D}} \bm{\psi} $ in $\xi \cdot \bm{D} \bm{\psi} $ and the remaining terms. In the second step we have regrouped various terms to factor out $\left( \xi \cdot \bm{e}^a \right)$. In the third step we have regrouped terms such that the coefficient of $\left( \xi \cdot \bm{e}^a \right)$ is the energy-momentum three-form $\bm{\cT}_a$  introduced in \ref{new_tau}.

Combining \ref{partI} and \ref{partII},  the Lorentz-diffeomorphism Noether current  for the Rarita-Schwinger part of the Lagrangian can be expressed as,
\bea
\bm{J}_\xi^{\rom{RS}}&=&\bm{\theta}(\bm{\psi},\cK_{\xi}\bm{\psi})-\xi\cdot \bm{L}
\nn
\\
&=& \frac{i}{2 \kappa^2} \bar{\bm{\psi}} \w \gamma_{*} \bm{\gamma} \w \hat{\bm{D}} (\xi \cdot \bm{\psi}) - \bm{\cS}^{cd} \left(  \half E_c \cdot E_d \cdot \bm{d \xi} + \xi \cdot \bm{K}_{cd} \right) \nn
\\
&& - \bm{\cT}_a \left( \xi \cdot \bm{e}^a \right) + \frac{i}{2 \kappa^2} \left( \overline{\xi \cdot \bm{\psi}} \right) \ \gamma_{*} \bm{\gamma} \w \hat{\bm{D}} \bm{\psi}. \label{RS_Noether}
\eea
The first term in the above equation can be rewritten as a sum of a total derivative and other terms (cf. \ref{RS_psi_var}),
\bea
\frac{i}{2 \kappa^2} \bar{\bm{\psi}} \w \gamma_{*} \bm{\gamma} \w \hat{\bm{D}} (\xi \cdot \bm{\psi}) &=& \bm{d} \left(\frac{i}{2 \kappa^2} \bar{\bm{\psi}} \w \gamma_{*} \bm{\gamma} \left(\xi \cdot \bm{\psi} \right) \right) - \frac{i}{2 \kappa^2} \bar{\bm{\psi}} \overleftarrow{\hat{\bm{D}}} \w \gamma_{*} \bm{\gamma} \left(\xi \cdot \bm{\psi} \right)
\nn
\\
&& + \frac{i}{2 \kappa^2} \bar{\bm{\psi}} \w \gamma_{*} \gamma_{a} \bm{T}^a \left( \xi \cdot \bm{\psi} \right),
\eea
while the last term in \ref{RS_Noether} can be Majorana flipped (cf. \ref{majorana_flip2}) to get,
\be
\frac{i}{2 \kappa^2} \left( \overline{\xi \cdot \bm{\psi}} \right) \ \gamma_{*} \bm{\gamma} \w \hat{\bm{D}} \bm{\psi} = -\frac{i}{2 \kappa^2} \bar{\bm{\psi}}\overleftarrow{\hat{\bm{D}}}\w \gamma_{*} \bm{\gamma} \left( \xi \cdot \bm{\psi} \right). 
\ee
We finally have,
\be
\bm{J}_\xi^{\rom{RS}}=\bm{d} \left[\frac{i}{2 \kappa^2} \bar{\bm{\psi}} \w \gamma_{*} \bm{\gamma} \left(\xi \cdot \bm{\psi} \right) \right] - \bm{\cT}_a \left( \xi \cdot \bm{e}^a \right) - \bm{\cE}_{\psi} \left(\xi \cdot \bm{\psi} \right) 
- \bm{\cS}^{cd} \left( \half E_c \cdot E_d \cdot \bm{d \xi} + \xi \cdot \bm{K}_{cd} \right).
\ee
The full Lorentz-diffeomorphism Noether current is therefore, 
\bea
\bm{J}_\xi &=& \bm{J}^{\rom{grav}}_\xi + \bm{J}^{\rom{RS}}_\xi \nn \\
&=&  \bm{dQ}^{\rm grav}_\xi +\bm{d}\left[\frac{i}{2\kappa^{2}}\bar{\bm{\psi}}\w \gamma_{*}\bm{\gamma} \left(\xi\cdot \bm{\psi}\right) \right] + (\bm{\mathcal{E}}_{a} - \bm{\Tau}_{a}) (\xi \cdot \bm{e}^{a})  +\bm{\cE}_{\bm{\psi}}\left(\xi\cdot \bm{\psi}\right) \nn \\
&& +\left( \bm{\mathcal{E}}^{cd} -\bm{\mathcal{S}}^{cd} \right) \left( \half E_c \cdot E_d \cdot \bm{d \xi} + \xi \cdot \bm{K}_{cd} \right).
\eea
On-shell we have $\bm{J}_\xi= \bm{d Q}_\xi $ where  the total Noether charge $\bm{Q}_\xi$ is 
\be
\bm{Q}_\xi=\bm{Q}^{\rm grav}_\xi+\bm{Q}^{\rm RS}_\xi,
\ee
with
\be
\bm{Q}^{\rm RS}_\xi  = \frac{i}{2\kappa^{2}}\bar{\bm{\psi}}\w \gamma_{*}\gamma_{a}\bm{e}^{a}\left(\xi\cdot \bm{\psi}\right), \label{Q_RS}
\ee
and where we recall that,
\bea
\bm{Q}^{\rm grav}_\xi 
&=& \frac{1}{32 \pi } \varepsilon_{ab}{}^{cd} \ \bm{e}^a \wedge \bm{e}^b  \left( \half E_c \cdot E_d \cdot \bm{d \xi} + \xi \cdot \bm{K}_{cd} \right) \\
&=&
\frac{1}{32 \pi } \varepsilon_{ab}{}^{cd} \ \bm{e}^a \wedge \bm{e}^b \left(\frac{1}{2}E_c \cdot E_d \cdot \left(d \bm{\xi} - \bm{T}^e (E_e \cdot \bm{\xi}) \right) - (E_e \cdot \bm{\xi}) (E_{[c} \cdot E^e \cdot \bm{T}_{d]}\right). \label{Q_grav}
\eea
Expressions \ref{new_theta}, \ref{new_tau}, and the expression for the Noether charge \ref{Q_RS} are the main results of this section. 

 \section{First law for supergravity}
\label{sec:first_law}
We now have all the ingredients necessary to formulate a first law for simple AdS supergravity. We recall that a first law is an identity relating the perturbed Hamiltonians for the horizon generating Killing field evaluated at the bifurcation surface and at spatial infinity.  The first variations of the Hamiltonians are constructed out of the Noether charge 2-form $\bm{Q}_\xi$ and symplectic potential 3-form  $\bm{\theta}$ as,
\be
\delta H_{\xi}=  \int_\Sigma {\bm \omega}(\varphi, \delta \varphi, \mathcal{K}_\xi \varphi) =  \int_{\partial \Sigma}\left(\delta \bm{Q}_{\xi}-\xi\cdot \bm{\theta}\right). \label{hamiltonians}
\ee
In this expression $\Sigma$ is a Cauchy surface in the spacetime and $\partial \Sigma$ is its boundary, $\bm{\omega}$ is the symplectic current 
\be
\bm{\omega}(\varphi, \delta_1 \varphi, \delta_2 \varphi) = \delta_1 \bm{\theta} (\varphi, \delta_2 \varphi) -  \delta_2 \bm{\theta} (\varphi, \delta_1 \varphi).
\ee
The Noether charge 2-form $\bm{Q}_\xi$ and symplectic potential 3-form  $\bm{\theta}$  for both the gravitational Lagrangian and the Rarita-Schwinger Lagrangian have been obtained in the previous sections. 
 
 A first step to discuss the first law is to specify the boundary conditions for the gravitational and Rarita-Schwinger fields. Asymptotically, we demand the metric to behave as global AdS,
\be
ds^{2}_{\rm AdS}=-\left(1+\frac{r^{2}}{L^{2}}\right)dt^{2}+\left(1+\frac{r^{2}}{L^{2}}\right)^{-1}dr^{2}+r^{2}\left(d\theta^{2}+\sin^{2}\theta d\phi^{2}\right).
\ee
A set of co-frames that capture the above metric is simply,
\begin{align}
\bm{e}^{0}_{\rm AdS} & = \left(1+\frac{r^{2}}{L^{2}}\right)^{1/2}dt, & \bm{e}^{1}_{\rm AdS} & = \left(1+\frac{r^{2}}{L^{2}}\right)^{-1/2}dr, \\
\bm{e}^{2}_{\rm AdS} & = r d\theta,&  \bm{e}^{3}_{\rm AdS} & = r \sin \theta d\phi.
\end{align}
The ``Dirichlet'' boundary conditions that define asymptotically AdS spacetimes in simple supergravity were worked out in \cite{Henneaux:1985tv}. They are, 
\bea
\bm{e}^{0} &=& \bm{e}^{0}_{\rm AdS} + \mathcal{O}(r^{-5}) dr +   \mathcal{O}(r^{-2}) dt +  \mathcal{O}(r^{-2}) d\theta +  \mathcal{O}(r^{-2}) d\phi, \\
\bm{e}^{1} &=& \bm{e}^{1}_{\rm AdS} + \mathcal{O}(r^{-4}) dr +   \mathcal{O}(r^{-3}) dt +  \mathcal{O}(r^{-3}) d\theta +  \mathcal{O}(r^{-3}) d\phi, \\
\bm{e}^{2} &=& \bm{e}^{2}_{\rm AdS} + \mathcal{O}(r^{-5}) dr +   \mathcal{O}(r^{-2}) dt +  \mathcal{O}(r^{-2}) d\theta +  \mathcal{O}(r^{-2}) d\phi, \\
\bm{e}^{3} &=& \bm{e}^{3}_{\rm AdS} + \mathcal{O}(r^{-5}) dr +   \mathcal{O}(r^{-2}) dt +  \mathcal{O}(r^{-2}) d\theta +  \mathcal{O}(r^{-2}) d\phi.
\eea
and 
\begin{align}
\psi_t &= (1-\gamma_1) \mathcal{O}(r^{-3/2}) ,  &
\psi_r &=(1+\gamma_1) \mathcal{O}(r^{-7/2}), \\
\psi_\theta &=(1-\gamma_1)  \mathcal{O}(r^{-3/2}), &
\psi_\phi &= (1-\gamma_1)  \mathcal{O}(r^{-3/2}).
\end{align}
These boundary conditions ensure that the symplectic current is finite at the boundary and that the symplectic flux through the boundary vanishes \cite{Hollands:2006zu}. The corresponding boundary conditions in the asymptotically flat setting were first discussed in \cite{Teitelboim:1977hc, Deser:1977hu}.

Let us assume that there is a stationary axisymmetric black hole solution to the field equations. Let the black hole horizon be a  bifurcate Killing horizon generated by the Killing field $k^{\mu}$. From general results on the bifurcate Killing horizon it follows that the Killing field  is a linear combination of time translation $t^{\mu}=(\partial/\partial t)^{\mu}$ and rotation $\phi^{\mu}=(\partial/\partial \phi)^{\mu}$,
\be
k^{\mu}=t^{\mu}+\Omega_H\phi^{\mu},
\ee 
where $\Omega_H$ is a constant representing horizon angular velocity and that  $k^{\mu}$ vanishes at the bifurcation 2-sphere.

Let us also assume that the Rarita-Schwinger field is smooth on the relevant part of the spacetime, i.e., in the neighbourhood of  the future and past horizons, at the bifurcation 2-sphere, and in the spacetime region outside the horizon all the way to infinity. Moreover, let us assume that the Rarita-Schwinger field is stationary and axisymmetric, i.e., 
\be
\mathcal{K}_{t}\bm{\psi}=0 \qquad \mathrm{and}  \qquad
\mathcal{K}_{\phi}\bm{\psi}=0 \qquad \implies  \qquad \mathcal{K}_{k}\bm{\psi}=0.
\ee
These conditions ensure stationarity and axi-symmetry of the solution to the field equations. Since the vector $k^{\mu}$ vanishes at the bifurcation 2-sphere,  the contribution from the $k \cdot \bm{\theta}$ term in the perturbed Hamiltonian
\be
\delta H_{k} =   \int_{\mathcal{B}}\left(\delta \bm{Q}_{k}-k \cdot \bm{\theta}\right),
\ee
is zero.  Our boundary conditions at infinity are such that there exists \cite{Hollands:2006zu} a 2-form $\bm{\Theta}$, such that
\be
\int_{\partial \Sigma_\infty} k\cdot \bm{\theta}= \delta \int_{\partial \Sigma_\infty} k\cdot \bm{\Theta}.
\ee
Hence, boundary Hamiltonians exist.

For the gravity sector, at the bifurcation 2-sphere, the contribution to the perturbed Hamiltonian $\delta H_k$ becomes $T_H \delta S$ .  Here $T_{H}=(\kappa_B/2\pi)$ and $S=(\textrm{A}/4)$, where $\kappa_B$ is the surface gravity and $A$ is the area of the bifurcation 2-sphere. This is because with $k^{\mu}=0$ on the bifurcation surface and $\delta k^{\mu}=0$ everywhere, one can argue that the variation of the temperature term is zero \cite{Iyer:1994ys,Prabhu:2015vua}. 

Since the Rarita-Schwinger field is assumed to be smooth on the horizon, the contribution of the Rarita-Schwinger field to the Noether charge \ref{Q_RS} at the bifurcation 2-sphere is zero: since $k^\mu=0$ at the bifurcation 2-sphere, $k\cdot \bm{\psi}$ vanishes. Hence on the bifurcation surface, the contribution from simple AdS supergravity is simply $T_{H}\delta S$.

At infinity, as is well known the contribution to the Hamiltonian from the  gravitational field yields the ADM mass $\mathcal{M}_{\rm ADM}$ and angular momentum $\mathcal{J}_{\rm ADM}$. Thus the variation of the gravitational Hamiltonian at infinity yields the variation of the ADM mass and angular momentum. While depending on the nature of the solution, the Hamiltonian for the Rarita-Schwinger field may or may not contribute at infinity. The boundary conditions we mentioned above are such that the \emph{supercharges} are finite. With these boundary conditions contributions to the energy and angular momentum from the Rarita-Schwinger field (which depends quadratically on the spinor field) vanish.  
Combining these elements, the stationary comparison version of the first law for black holes with bifurcate Killing horizons in simple AdS supergravity takes the form,
\be
T_H \delta S=\delta \mathcal{M}_{\rm ADM} -\Omega_{H} \mathcal{J}_{\rm ADM}.
\ee

In summary, we found that smooth,  stationary, axisymmetric Rarita-Schwinger field does not explicitly contribute to the black hole entropy. The extra term in the Noether charge vanishes at the bifurcation  surface. Near infinity, 
Rarita-Schwinger field falls-off sufficiently fast that it does not contribute to the integrals for the energy and angular momentum. Thus, the first law of  black hole mechanics in simple supergravity retains the same form as in pure general relativity.

\section{Conclusions}
\label{sec:discussion}
In this work we have applied and appropriately adapted the Lorentz-diffeomorphism Noether charge formalism of references \cite{Jacobson:2015uqa, Prabhu:2015vua} to four-dimensional gravity coupled to a Majorana field and to a Rarita-Schwinger field. In section \ref{sec:Majorana} we  studied the Lorentz-diffeomorphism Noether charge for a spin-1/2 Majorana field.  The Majorana condition brings in certain new elements in the computation. It served as a warm-up for the Rarita-Schwinger field in the context of  simple supergravity considered in section \ref{sec:supergravity}. As we saw in that section the Majorana nature of the Rarita-Schwinger field played an important role in the computations. A key result of our work is expression \ref{Q_RS} for the contribution to the Lorentz-diffeomorphism Noether charge due to the Rarita-Schwinger field. Using this Noether charge we formulated a stationary comparison version of the first law in section  \ref{sec:first_law}. 

In our analysis of the first law with the Rarita-Schwinger field we made two important assumptions: (i)  The Rarita-Schwinger field is smooth everywhere in the region of interest, (ii) The Rarita-Schwinger field is annihilated by the Kosmann derivative with respect the horizon generating Killing field. Using these assumptions, we concluded that the Rarita-Schwinger field does not contribute to the first law at the bifurcation surface. Perhaps these assumptions are too restrictive. This situation should be compared to the analysis of the Yang-Mills field by Sudarsky and Wald \cite{Sudarsky:1992ty, Sudarsky:1993kh}. Under similar assumptions, namely (i) a smooth Yang-Mills field can be chosen on the spacetime, and (ii) it is annihilated by the Lie derivative with respect the horizon generating Killing field, they also concluded that the Yang-Mills field does not contribute to the first law at the bifurcation surface. Over the years, this conclusion has been refined. In 2003 Gao \cite{Gao:2003ys} argued that the Yang-Mills field does contribute to the first law at the horizon, but he was not able to write the contribution as a potential times the perturbed charge without making additional assumptions. In 2015 Prabhu \cite{Prabhu:2015vua} by formulating the problem in terms of the principal bundle gave a satisfactory discussion of the first law for gravity coupled to a  Yang-Mills field. He showed that the Yang-Mills field contributes to the first law both at the bifurcation surface and at infinity. The contributions are of the form potential times the perturbed charge, and generically it is not possible to write the two terms as the `difference in the potential between infinity and the bifurcation surface' times the perturbed charge.

It is natural to speculate that similar refinements are to be found with the Rarita-Schwinger field. A reason our analysis is ill-equipped to address this question is that we have ignored the fermionic gauge symmetry of the Rarita-Schwinger field. The fermionic gauge symmetry is supersymmetry---a spacetime symmetry. It changes the frame field as well. The principal bundle formalism of Prabhu \cite{Prabhu:2015vua}, though quite general, is not equipped to handle supersymmetry. Perhaps a formulation of the first law is possible using the superspace formalism of supergravity. The superspace  is discussed at length in the mathematical physics literature \cite{Castellani:1991et, Castellani:1991eu,DeWitt:1992cy, Buchbinder:1995uq}. In such a formulation, we expect that the above mentioned shortcomings can be addressed and that supercharges may feature in the first law. Such a discussion would shed further light on black holes in supergravity. We leave this for future work.

For a class of \emph{supersymmetric} black holes it is known that smooth (at least at the future horizon) normalisable \emph{linearised} fermionic hair modes exist \cite{Aichelburg:1982pi, Jatkar:2009yd}. It will be very interesting to understand how these modes appear in the first law for the corresponding black holes. 

We hope to return to these questions in our future work.

\subsection*{Acknowledgements} We thank Alok Laddha, Bindusar Sahoo, and Simone Speziale for discussions. SC thanks AEI Potsdam and CMI Chennai for hospitality towards the initial and final stages of this project respectively. The work of PBA, SkJH, and AV  is supported in part by the Max Planck Partnergroup ``Quantum Black Holes'' between CMI Chennai and AEI Potsdam and by a grant to CMI from the Infosys Foundation. The research of SkJH is also supported in part by the Czech Science Foundation Grant 19-01850S.
  The work of SC is supported in part by the INSPIRE Faculty fellowship (Reg. No.  DST/INSPIRE/04/2018/000893).

\end{document}